\documentclass[aps,showpacs,preprint,nofootinbib]{revtex4}

\begin{document}
\author{N. Kiriushcheva$^1$ and S. V. Kuzmin$^{2}$\\ 
{\it $^1$Department of Applied Mathematics,
University of Western Ontario,}\\
{\it London, Ontario, N6A 5B7 Canada}\\
{\it $^2$Faculty of Arts and Social Science,
Huron University College,}\\
{\it 1349 Western Road, London, Ontario, N6G 1H3 Canada}
\\email: nkiriush@uwo.ca, skuzmin@uwo.ca}

\title{Comments on ``A note on first-order formalism and 
odd-derivative actions'' by S.~Deser}  
\date{\today}

\begin{abstract}
We argue that the obstacles to having a first-order formalism for 
odd-derivative actions presented in a pedagogical note by Deser are based on 
examples which are not 
first-order forms of the original actions. The 
general derivation of an equivalent first-order form of the original 
second-order action is illustrated using the example of topologically massive
electrodynamics (TME). The correct first-order formulations of the TME model 
keep intact the gauge invariance presented in its second-order form
demonstrating that the gauge invariance is not 
lost in the Ostrogradsky process.  

\end{abstract}

\maketitle

In a recent pedagogical note \cite{Deser}, S. Deser discusses obstacles to the
first-order formulations of theories with an odd number of derivatives 
occurring in the  actions, arguing that
first-order formulations might destroy gauge invariance. The title of
the note is very general and creates impression that there is a problem
in applying first-order formulations to all odd-derivative 
actions, which we argue below to be incorrect. The simplest counter-example is 
a pure Yang-Mills
action which contains odd-derivative terms, but its first- and second-order 
formulations are equivalent and both of them are gauge invariant. 

The simplest illustration of an
odd-derivative action in \cite{Deser} is topologically massive 
electrodynamics (TME) in 3D, where the Lagrangian 
in the standard 
second-order form is \footnote{The metric convention that we use 
is $\eta^{\mu \nu} =
diag\left(-,+,+,...\right)$, 
$\epsilon ^{\mu \nu \alpha ...}$ is totally antisymmetric tensor, 
$\epsilon _{\mu \nu \alpha ...}=-\epsilon ^{\mu \nu \alpha ...}$, 
$\epsilon ^{012...}=1$.}

\begin{equation}
\label{1}M_{2}(A)= -\frac{1}{4}f_{\mu \nu}f^{\mu \nu}-\frac{m}{2}
\epsilon^{\mu \nu \alpha}f_{\mu \nu} A_\alpha
\end{equation}
where $A_\mu$ is the abelian vector field and

\begin{equation}
\label{2}f_{\mu \nu}=\partial_\mu A_\nu - \partial_\nu A_\mu,
\end{equation}
is just a short notation, not an independent variable. 
This sum of the Maxwell Lagrangian and a Chern-Simons term has been 
proposed in three dimensions 
and exhaustively analyzed in its second-order form in
\cite{DJT}. (We follow the
notation of \cite{Deser} which is slightly different from that of \cite{DJT}.) 

The first-order formulations of (\ref{1}) are discussed in \cite{Deser}. 
However, neither of the proposed first-order formulations
of (\ref{1}) reproduces the equations of motion 
that follow from (\ref{1})

\begin{equation}
\label{3} \frac{\delta}{\delta A_\alpha}M_2 = \partial_\mu f^{\mu \alpha}-
m \epsilon^{\mu \nu \alpha} f_{\mu \nu}=0.
\end{equation}
and thus should not be used as grounds for any comparison between the first and
second-order forms and, in particular, 
to reach any conclusion about the loss of gauge invariance in this 
transition.

In his introduction, the author of \cite{Deser} refers to a first-order form 
as a particular case of
the general Ostrogradsky procedure for lowering the number of derivatives 
appearing by adding 
new variables.\footnote{This definition, in our opinion, is the most complete 
and the correct one to be used in field theory. However, to the best of our 
knowledge, neither Ostrogradsky \cite{Ostr} 
nor authors of further modifications of his methods (e.g. \cite{GLT} and 
\cite{DD}) followed this definition, and only lowering
order of temporal derivatives was considered.} 
This general procedure in which new variables are added is a well 
defined operation that leads to an equivalent variational problem of the 
lower order Lagrangian through  using extra fields as Lagrange multipliers 
and, as we will 
demonstrate, does not destroy gauge invariance. 
In the case of (\ref{1}) this procedure gives the first-order Lagrangian
\begin{equation}
\label{4}M_{1}(A,F,W)=-\frac{1}{4}F_{\mu \nu}F^{\mu \nu} - \frac{m}{2}
\epsilon^{\mu \nu \alpha}F_{\mu \nu}A_\alpha + \frac{1}{2}W^{\mu \nu}
\left(F_{\mu \nu} - f_{\mu \nu}\right).
\end{equation}  

In (\ref{4}) there are two new independent antisymmetric tensor fields: 
$F_{\mu \nu}, W^{\mu \nu}$. The
coefficient in front of the last term is chosen so as to yield the   
conventional first-order form in the limit of $m=0$ (i.e. the first-order 
formulation of the Maxwell Lagrangian).
Let us demonstrate the equivalence of (\ref{1}) and (\ref{4}). 
Performing a variation of (\ref{4}) with respect to all 
independent variables we obtain the system of equations

\begin{equation}
\label{5}\frac{\delta}{\delta F_{\mu \nu}}M_1=-\frac{1}{2}F^{\mu \nu}-
\frac{m}{2}\epsilon^{\mu \nu \alpha}A_\alpha + \frac{1}{2}
W^{\mu \nu} = 0,
\end{equation}      
\begin{equation}
\label{6}\frac{\delta}{\delta W^{\mu \nu}} M_1 = \frac{1}{2}F_{\mu \nu} -
\frac{1}{2}f_{\mu \nu} = 0,
\end{equation}
\begin{equation}
\label{7}\frac{\delta}{\delta A_\rho} M_1= -\frac{m}{2}\epsilon^{\mu \nu \rho}
F_{\mu \nu} + \partial_{\mu} W^{\mu \rho} = 0.
\end{equation}

Solving (\ref{5},\ref{6}) for $F^{\mu \nu}$ and $W^{\mu \nu}$ and 
substituting them into (\ref{7}) we recover the 
original equation of motion (\ref{3}). Moreover, upon 
substituting into (\ref{4}) we recover the original action (\ref{1}), 
thus proving equivalence of (\ref{4}) with (\ref{1}). (We have done this 
for pure pedagogical reasons because reducing the number of
derivatives by using extra fields and Lagrange multipliers always gives
an equivalent formulation). For the TME model there is the possibility of an 
even 
simpler formulation, which is the result of the particular form of 
(\ref{5}) which can be solved algebraically for the field 
$F^{\mu \nu}$. Substitution of this solution into (\ref{4}) produces an 
equivalent form of the Lagrangian in which fewer variables appear
\footnote{The identity $\epsilon^{\mu \nu \alpha}\epsilon_{\mu \nu}^
{\hspace{2mm} \beta}=-2\eta ^{\alpha \beta}$ was used.}
\begin{equation}
\label{8}\tilde M_1(A,W)= \frac{1}{4}W_{\mu \nu}W^{\mu \nu} - 
\frac{1}{2}W_{\mu \nu}
f^{\mu \nu} - \frac{m}{2}\epsilon^{\mu \nu \alpha}W_{\mu \nu}A_\alpha -
\frac{m^2}{2}A_\mu A^\mu.
\end{equation}
This is also a general result: if an
algebraic equation of motion (such as (\ref{5})) can be solved for a field, 
this field can be eliminated from the Lagrangian.\footnote{Not 
all actions, especially with higher-order derivatives, satisfy
such a condition and it is better to start from an unambiguous form such as 
(\ref{4}). 
Also, in the case of non-linear theories, this longer way, may even be
preferable, because a
term with derivatives will be present only in terms bilinear in the fields and 
all 
non-linearity becomes purely algebraic (i.e. without derivatives).} As before, 
one can 
repeat the calculations of eqs. (\ref{4}-\ref{7}) and show explicitly 
that elimination of $W_{\mu \nu}$ from the equations of motion
\begin{equation}
\label{8a}\frac{\delta}{\delta W_{\mu \nu}} \tilde M_1 =- \frac{1}{2}
W^{\mu \nu} 
-\frac{1}{2}f^{\mu \nu}-\frac{m}{2}\epsilon^{\mu \nu \rho}A_{\rho} = 0,
\end{equation}
\begin{equation}
\label{8b}\frac{\delta}{\delta A_\rho} \tilde M_1= \partial_{\mu} W^{\mu \rho}
-\frac{m}{2}\epsilon^{\mu \nu \rho}W_{\mu \nu} -m^{2}A^{\rho}  = 0
\end{equation}
which can be used to recover both eqs. (\ref{3}) and (\ref{1}).

The actions of eqs. (\ref{4}) and (\ref{8}) are both true first-order 
formulations of (\ref{1}) 
and these actions, not the actions considered in \cite{Deser}, lead to the 
equations of motion (\ref{3}) which arise in the second-order form of TME 
\cite{DJT}.

Having true first-order formulations, we can check whether there is indeed 
a loss of gauge invariance in first-order formulations and see if this  
obstacle is an artifact of working with a wrong formulation of first-order 
TME. It would seem  unreasonable to
expect a loss of gauge invariance for formulations that lead to the same 
equations of motion (\ref{3}), which are invariant under the abelian gauge 
transformation

\begin{equation}
\label{9}\delta A_\mu = \partial_\mu \phi.
\end{equation}

Of course, there is a difference between the TME model and ordinary gauge 
theories such as Maxwell or Yang-Mills (YM). The last two are invariant 
{\it exactly} under the corresponding gauge transformations in both first- 
and second-order formulations, whereas the second-order
TME is invariant {\it only up to a surface term}. Variation of (\ref{1}) 
leads to the expression

\begin{equation}
\label{10}\delta M_2(A) = -\frac{m}{2}\epsilon^{\mu \nu \alpha}f_{\mu \nu}
\partial_\alpha \phi = -m \epsilon^{\mu \nu \alpha} \partial_\mu A_\nu 
\partial_\alpha \phi
\end{equation} 
that, by addition of zero in the form $0 \equiv -m\epsilon^{\mu \nu \alpha}
\partial_\mu\partial_\alpha A_\nu \hspace{1mm} \phi$, 
can be presented as a surface term \footnote{Alternatively, a different zero,
$0 \equiv -m\epsilon^{\mu \nu \alpha} A_\nu 
\partial_\mu \partial_\alpha \phi$, can be added; this gives a different 
surface term $\delta M_2(A) = - \partial_\mu \left(m \epsilon^{\mu \nu \alpha} 
A_\nu \partial_\alpha \phi\right)$.}
\begin{equation}
\label{11}\delta M_2(A) = -\partial_\alpha \left(m \epsilon^{\mu \nu \alpha} 
\partial_\mu A_\nu \hspace{1mm} \phi \right).
\end{equation}

An invariance up to a surface term requires us to impose a condition
on the asymptotic behavior of gauge parameters, whereas in the case of an 
exact gauge invariance (for such Lagrangians as Maxwell and Yang-Mills)  
gauge parameters are free from such restrictions.
The consequences of this difference is an interesting question to discuss but 
it is beyond the scope of our short comment. This
``up to surface term invariance'' is also
present in the first-order formulations of TME.

It is not difficult to check that the Lagrangian (\ref{4}) is invariant under 
transformations

\begin{equation}
\label{12}\delta A_\mu = \partial_\mu \phi, \hspace{2mm} \delta F_{\mu \nu} 
= 0, \hspace{2mm}
\delta W^{\mu \nu}=m \epsilon^{\mu \nu \alpha}\partial_\alpha \phi
\end{equation}
and (\ref{8}) under
\begin{equation}
\label{13}\delta A_\mu = \partial_\mu \phi, \hspace{2mm}
\delta W^{\mu \nu}=m \epsilon^{\mu \nu \alpha}\partial_\alpha \phi.
\end{equation}

The Lagrangians in both first-order formulations are invariant up to a 
surface term which is the same surface term as the one appearing
in the second-order formulation. Moreover, all equations of
motion (\ref{5}-\ref{7}) and (\ref{8a},\ref{8b}) are exactly invariant  
under (\ref{12}) and (\ref{13}), in full agreement with 
the results of the second-order formulation,
thus demonstrating that the presence of the Chern-Simons term does not
create any ``fundamental distinction'' between the first- and second-order 
formulations of the model.
A detailed canonical analysis of the first-order formulation of TME using 
the Dirac constraint formalism including a derivation of the transformations 
of Eqs. (\ref{12},\ref{13}) is the subject of a forthcoming paper \cite{GKKM}.

The second example considered in \cite{Deser} is the ``first-order form'' of 
3D topologically massive gravity (TMG) \cite{DJT}. In this case the 
combination 
(Eqs. (5b+7a) of \cite{Deser}) used by the author is again not a first-order 
formulation of the original
model. The equations of motion for these 
formulations of \cite{Deser} are different because affinity, according to 
\cite{Deser}, is no longer a metric 
one in the so-called ``fully Palatini model''.\footnote{Which is, we believe, 
equivalent in the author terminology to ``first-order formulation''.}
The true first-order formulation 
of this model requires more involved calculations compared to the TME model, 
in particular, because the derivatives of the metric up to third order are 
present in the original action and this means that more additional fields are
required to convert 
it into an equivalent first-order form. 
Such calculations would be another illustration of the general procedure for 
lowering the order of 
derivatives appearing in the action by means of Lagrange multipliers.
\\
\\
{\large \bf Acknowledgments}\\

The authors are grateful to R. N. Ghalati, D. G. C. McKeon,  
and A. Zvelindovsky for 
discussions and reading the final version of the manuscript.

\end{document}